\documentclass[preprint,showpacs,showkeys,superscriptaddress]{revtex4}
\usepackage{graphicx}
\usepackage{amsmath, amsthm, amsfonts,amssymb}

\begin{document}

\title{Statistical Properties of Cross-Correlation in the Korean Stock Market}

\author{Gabjin Oh}
\affiliation{Division of Business
Administration, Chosun University, Gwangju 501-759, Republic of
Korea}

\author{Cheoljun Eom}
\affiliation{Division of Business Administration, Pusan National
University, Busan 609-735, Republic of Korea}

\author{Fengzhong Wang}
\affiliation{Center for Polymer Studies and Department of Physics,
Boston university, Boston, MA 02215, USA}

\author{Woo-Sung Jung}
\email{wsjung@postech.ac.kr} 
\affiliation{Center for Polymer Studies and Department of Physics,
Boston university, Boston, MA 02215, USA}
\affiliation{Graduate Program for Technology and Innovation Managemant,
Pohang University of Science and Technology, Pohang 790-784, Republic of Korea}
\affiliation{Department of Physics, Pohang University of
Science and Technology, Pohang 790-784, Republic of Korea}

\author{H. Eugene Stanley}
\affiliation{Center for Polymer Studies and Department of Physics,
Boston university, Boston, MA 02215, USA}

\author{Seunghwan Kim}
\affiliation{Department of Physics, Pohang University of
Science and Technology, Pohang 790-784, Republic of Korea}
\affiliation{Asia Pacific Center for Theoretical Physics, Pohang 790-784, Republic of Korea}

\begin{abstract}

We investigate the statistical properties of the correlation
matrix between individual stocks traded in the Korean stock market
using the random matrix theory (RMT) and observe how these affect
the portfolio weights in the Markowitz portfolio theory. We find
that the distribution of the correlation matrix is positively
skewed and changes over time. We find that the eigenvalue
distribution of original correlation matrix deviates from the
eigenvalues predicted by the RMT, and the largest eigenvalue is 52
times larger than the maximum value among the eigenvalues
predicted by the RMT. The $\beta_{473}$ coefficient, which reflect
the largest eigenvalue property, is 0.8, while one of the
eigenvalues in the RMT is approximately zero. Notably, we show
that the entropy function $E(\sigma)$ with the portfolio risk
$\sigma$ for the original and filtered correlation matrices are
consistent with a power-law function, $E(\sigma) \sim
\sigma^{-\gamma}$, with the exponent $\gamma \sim 2.92$ and those for Asian currency crisis decreases significantly.

\end{abstract}

 \maketitle

\section{Introduction}
Financial markets have been known as representative complex
systems, which are organized by various unexpected phenomenon
according to non-trivial interactions among heterogeneous agents
\cite{Thomas Lux}. The study of complex economic systems is not
easy because we do not know the control parameters that govern
economic systems and can not easily apply the parameters we do
know to economic systems. However, much research has been
conducted to understand the statistical properties of financial
time series \cite{Mantegna(a), Oh}. In particular, the analysis of
financial data by various methods developed in statistical physics
has become a very interesting research area for physicists and
economists \cite{Mantegna&Bouchaud}. There is practical
\cite{Elton, Okhrin, Sornette} as well as scientifically important
value in analyzing the correlation coefficient between stock
return time series because this contains a significant amout of
information on the nonlinear interactions in the financial market
and is a parameter in terms of the Markowitz portfolio theory. The
correlation matrix between stocks, which has unexpected properties
due to complex behaviors, such as temporal non-equilibrium,
mispricing, bubbles, market crashes and so on, is an important
parameter to understand the interactions in the financial market
\cite{Noh}. To analyze the correlation matrix, previous studies
presented various statistical methods, such as principal component
analysis (PCA) \cite{PCA}, singular value decomposition (SVD)
\cite{SVD} and factor analysis (FA) \cite{FA}. Here, to analyze
the actual correlation matrix, we employ the random matrix theory
(RMT), which was introduced by Wigner, Dyson and Metha
\cite{MehtaB, Wigner, Dyson, Mehta}. It can explain the
statistical properties of energy levels in complex nuclei well.
The RMT method is a useful method for eliminating the randomness
in the actual correlation matrix \cite{Guhr1, Sengupta, Seba,
Utsugi, Guhr2, Shari}. Recently, Laloux {\em et al} (1999)
\cite{Laloux} and Plerou {\em et al} (1999) \cite{Plerou1}
analyzed the correlation matrix of financial time series by the
RMT method. The authors found that 94\% of the eigenvalues of
correlation matrix can be predicted by the RMT, while the other
6\% of the eigenvalues deviated from the RMT. In addition, Plerou
et al (2002) \cite{Plerou2} applied the RMT method to a United
States stock market and observed that the correlation matrix of
stock markets consist of random and non-random parts, which have a
useful information in the financial market. The eigenvector
deviations from the RMT show a very stable state over a whole
period. We investigate the various statistical properties of the
correlation matrix of 473 daily stock return time series traded in
the Korean stock market from 1 January 1993 to 31 May 2003. We
find that the distribution of the correlation matrix is positively
skewed and changes over the whole time. Using the RMT method, we
show that the correlation matrix contains meaningful information
as well as random property. Notably, we show that for both the
original, $C_{original}$, and filtered correlation, $C_{filter}$,
matrices the entropy function, $E(\sigma)$, with the portfolio
risk, $\sigma$, is consistent with a power-law function,
$E(\sigma) \sim \sigma^{-\gamma}$, with an exponent $\gamma \sim
2.92$. In the following section, we describe the data and methods
used in this paper. In Section 3, we present the verification
results. Finally, we end with a conclusion.

\section{Data and Method}

In this paper, we investigate the statistical properties of the
correlation matrix of the 473 daily stock returns traded on the
Korean stock market from 3 January 1993 to 31 May 2003. The data
obtained from the Korea Stock Exchanges cover 2845 days. To
understand the non-trivial interactions, we calculated the
correlation matrix between stocks for the whole period as well as
sub-periods by shifting 21 days with 250 data points. We propose a
verification process to analyze the statistical properties of the
correlation matrix between stock returns. First, we estimated the
statistical properties of the correlation matrices using the RMT
method. Second, we calculate the entropy of the portfolio weights
using the Markowitz portfolio theory. Before demonstrating the
verification process, we introduces the RMT, which was proposed by
Wigner, Dyson, and Metha, {\em et al.} and Markowitz portfolio
theory (MPT) \cite{Markowitz} introduced by Markowitz in 1952. We
created $N$ (number of company) data sets with $L$ data points
following $iid(0,1)$. Let the created data be denoted by the
symbol $G$. Here, the $G$ is a matrix ($N\times L$) with the
random elements and the correlation matrix is defined by

\begin{equation}\label{e1}
C_{random} = \frac {1}{L}GG^T,
\end{equation}
where $G^T$ is the transpose of $G$, and the correlation between
elements is approximately zero. If $N \to \infty$ and $L \to
\infty$, the eigenvalue spectrum of RMT is calculated by using

\begin{equation}\label{e2}
P_{random}(\lambda) = \frac {Q}{2\pi} \frac{\sqrt{(\lambda_{+} -
\lambda)(\lambda - \lambda_{-})} }{\lambda},
\end{equation}
where the eigenvalues $\lambda$ lie within $\lambda_{-} \leq
\lambda \leq \lambda_{+}$, $Q \equiv \frac {L}{N}$, and the
maximum and minimum eigenvalue of RMT, $C_{random}$, are given by

\begin{equation}\label{e3}
\lambda \pm \equiv 1 + \frac{1}{Q} \pm 2\sqrt{\frac{1}{Q}}.
\end{equation}

If $L$ and $N$ have a limitable length, then the eigenvalue
spectrum shows gradual decrease from the theoretical values of the
largest eigenvalue predicted by the RMT.

We next explain the MPT to select the optimal portfolio sets among
all stocks. The MPT method introduced by Markowitz in 1952 is
generically known as the mean-variance theory. The purpose of MPT
is to minimize the portfolio risk in a given portfolio return,
which can be quantified by the variance and defined as follows.

\begin{equation}\label{e4}
\Omega = \sum_{i=1}^{N}\sum_{j=1}^{N} \omega_{i} \omega_{j} C_{ij}
\sigma_{i}\sigma_{j},
\end{equation}
where $\omega_{i}$ is the portfolio weight of stock $i$, which can
be calculated using two Lagrange multipliers, $\sigma_{i}$ is the
standard deviation of stock $i$, and $C_{ij}$ is the correlation
coefficient between stock $i$ and stock $j$. 
In this work, we use the no short-selling constraint for portfolio weights \cite{Mantegna&Bouchaud}, i.e. we assume that all the weights are non negative numbers ($\omega_i>0$, $\forall$ i=1,$\dots$,N). We also normalize portfolio weights in such a way that $\sum_{i=1}^{N}\omega_i=1$. The portfolio return, $\mu$, also is
calculated by

\begin{equation}\label{e5}
\mu = \sum_{i=1}^{N}\omega_{i}\mu_{i},
\end{equation}
where $\mu_{i}$ is the mean value of stock $i$. We next considered
the portfolio weights because these could determine the portfolio
efficiency frontier lines. We used Shannon's entropy method to
quantify the statistical properties of the portfolio weights since $\sum_{i=1}^{N} \omega_{i} = 1$,
defined by
\begin{equation}\label{e6}
E = \sum_{i=1}^{N} -P_{i}\ln(P_{i}),
\end{equation}
where $P_{i}$ is the portfolio weight $w_{i}$.

Using the eigenvalue distribution predicted by the equation 2, we
estimated a random part from the original correlation matrix and
as the previous paper \cite{Plerou2}, divided it two parts as
follows.

\begin{equation}\label{e7}
C_{original} = C_{random} + C_{filter}.
\end{equation}

Based on how many random elements existed in the correlation
matrix, we analyzed the non-trivial interactions between stocks.
In addition, to estimate the eigenvalue properties, we created the
data sets by using each eigenvector element.

\begin{equation}\label{e8}
R(t) \equiv \sum_{i=1}^{N}V_{i}r_{i}(t),
\end{equation}

where $r_{i}(t)$ is the $i\-th$ stock return at time $t$, and
$V_{i}$ is the $i\-th$ eigenvector. To observe the eigenvalue
properties divided by the RMT method, we created the data sets,
$R^{Random}(t)$ and $R^{Largest}(t)$, reflecting the eigenvalue
properties of both $C_{random}$ and $C_{filter}$, respectively,
and, by the one-factor model, widely acknowledged in the financial
literature as a pricing model, we calculated the relationship
between the created time series and the market factor, which
influences all stocks in the market and is defined by

\begin{equation}\label{e9}
r_{i}(t) = \alpha_i +\beta_{i}R_{Market}(t)+\epsilon_{i}(t),
\end{equation}
where $R_{Market}$ is the KOSPI market index, $\alpha_{i}$ and
$\beta_{i}$ are the regression coefficients of stock $i$ and use
the $\beta$ coefficient as the measurement to quantify the
relationship between created data sets and market index.

\section{Results}
\label{sec:RESULTS}

In this section, we analyze the various statistical features of
the correlation matrix of 473 daily stock returns listed on the
Korean stock markets from 3 January 1993 to 31 May. 2003 using the
random matrix theory and Markowitz portfolio theory. We present
the results on the statistical properties of the correlation
matrix, such as its distribution, eigenvalue spectrum and entropy
of portfolio weights calculated by MPT. Fig. 1(a) and (b) show the
distribution of the correlation matrices of the original and
random data sets. Fig. 1(c) shows the distribution of correlation
matrices calculated by shifting 21 days with 250 data points. Fig.
1(d) displays the average value of correlation matrices of Fig.
1(c). In Fig. 1(a), we find that the distribution of the
correlation matrix between stocks for a whole period is positively
skewed and shows a significant difference from that for the random
interaction in Fig. 1(b). In Fig. 1(c), we show that the
distribution of the correlation matrix changes considerably over
the whole time. Especially, in Fig. 1(d), during the Asian
currency crisis, the mean values of the correlation coefficients
significantly increased. In other words, the dynamically changes
were caused by the complex behavior of the market crash, unlike
the case of random interactions. Our findings confirm that all the
possible interactions in the Korean stock market deviated from
those for the random interaction.

We next decompose the original correlation matrix into the random
$C_{random}$ and filter $C_{filter}$ parts using the RMT method to
extract the meaningful information from the original correlation
matrix. Fig. 2 shows the eigenvalue distribution of the
correlation matrix in the Korean stock market. In Fig. 2, the
solid-line (orange) is the eigenvalue spectrum predicted by the
RMT, and the red circles and blue circles indicate the eigenvalue
distributions of the original time series and random data sets,
respectively. In Fig. 2, we find that the eigenvalue distribution
of the RMT method is very similar to one from the random data,
while that for the real time series significantly shows different
behavior. Moreover, the largest eigenvalue is 52 times larger than
the largest eigenvalue of the RMT. The large values are greater
than 25 times those in the United States stock market
\cite{Plerou2}.

To characterize the statistical properties of each eigenvalues, we
created the return time series using equation \ref{e8} and
calculated the slopes $\beta$ between those and the KOSPI market
index using equation \ref{e9}. Fig. 3(a) and (b) shows the
distribution of the eigenvector elements corresponding to both the
largest eigenvalue, $\lambda_{473}$ and $\lambda_{100}$, one of
eigenvalues of the RMT, respectively. Fig. 3(c) and (d) show the
$\beta$ coefficient between the KOSPI market index and the time
series created. We find that the $\beta_{473}$ between the market
index and time series is 0.8, while one from the time series
created using the eigenvector elements predicted by the RMT is
approximately zero. We argue that the largest eigenvalue can
explain the market properties well, but one from the ranges
predicted by the RMT is uncorrelated to the market index. We also
decomposed the original correlation matrix according to each
eigenvalue divided by the RMT method. Fig. 4 shows the
distribution of various correlation matrices. The red circles,
blue diamonds, black squares and pink triangles indicate the
correlation matrices of the original, random, filter and largest
eigenvalues, respectively. Through the above findings, we can
expect that the distribution of the random correlation matrix
$C_{random}$ follows a Gaussian distribution, while the
correlation matrix $C_{filter}$ estimated after removing the
random components from the original correlation matrix by the RMT
method has a similar distribution as the original time series. We
found that the correlation matrix reflecting the largest
eigenvalue property has an obvious difference from that of the
original time series.

To apply the RMT method to a portfolio optimization problem, we
analyzed the portfolio weights estimated by the MPT through
various correlation matrices. The important parameters are the
return, $\mu_{i}$, standard deviation, $\sigma$ and correlation
matrix, $C_{ij}$, of the original stock returns, which are needed
to calculate the portfolio weights of each stock. To calculate the
effects of the correlation matrix among the input parameters, we
apply the correlation matrices, $C_{filter}$, and $C_{random}$
divided by the RMT method. Fig. 5(a) shows the efficient portfolio
lines created using the various correlation matrices, such as
$C_{original}$, $C_{random}$, and $C_{filter}$. In Fig. 5 (a), we
found that the efficient frontier lines calculated with both the
original $C_{original}$ and filtered correlation matrices
$C_{filter}$ show very similar behavior, while that of the random
correlation matrix $C_{random}$ shows significant difference from
the original correlation. In addition, the efficient portfolio
frontier line of the random correlation matrix $C_{random}$ at a
given portfolio risk $\sigma$ overestimates the portfolio return,
$\mu$, by a greater amount than one of the original correlation
matrix. We next calculated the entropy of the portfolio weights
with each correlation matrix, such as $C_{original}$, $C_{filter}$
and $C_{random}$. Fig. 5(b) shows the relationship between the
portfolio risk, $\sigma$, and the entropy of the portfolio weights
for three types of correlation matrices according to a log-log
plot. We found that the entropy($\sigma$) for both the original
and filtered correlation matrices was approximately consistent
with a power-law function, $E(\sigma) \sim \sigma^{-\gamma}$ with
the exponent $\gamma \sim 2.92$, while there is no the power-law function in the relationship between the entropy and the portfolio return, $\mu$ and presented in Fig 5(c). We also calculated the exponents for each sub-periods by shifting 20 days with 500 data points to verify the stability over time the result observed in Fig. 5. We find that while the relationship between entropy of each portfolio weight and portfolio risk follow a power-law function, the exponent values, $\gamma$, calculated from each sub-periods changes over time and lie within $1.19 \leq \gamma \leq 3.23$. Especially, the $\gamma$ value calculated during the Asian currency crisis decreases significantly.

\section{Conclusions}
\label{sec:CONCLUSIONS}

We investigated the statistical properties of the correlation
matrix between the return time series of individual stocks traded
in the Korean stock market using the RMT method and observed the
effect of the correlation matrix applied to the Markowitz
portfolio theory. We found that the distribution of the
correlation matrix between stocks showed a positive skew and
dynamically changed over time. We found that the eigenvalue
distribution of the correlation matrix deviated from those of the
RMT, and the largest eigenvalue was 52 times larger than the
eigenvalues predicted by the RMT. The slopes $\beta$ between
market index and the time series corresponding to the largest
eigenvalue were 0.8, while those for the RMT were approximately
zero. Notably, we found that the entropy function $E(\sigma)$ of
portfolio weights with the portfolio risk $\sigma$ was consistent
with a power-law function, $E(\sigma) \sim \sigma^{-\gamma}$, with
the exponent $\gamma \sim 2.92$, while the relationship between the entropy and portfolio return $\mu$ is not a power-law function. We find that while for all sub-periods the exponents calculated from the relationship between entropy of each portfolio weight and portfolio risk follow a power-law function, those for sub-periods changed over time and lie within $1.19 \leq \gamma \leq 3.23$. Especially, the exponent $\gamma$ decreases significantly during Asian currency crisis. In the next step, we must
rigorously study the portfolio weights of other stock markets
because these play an important role in terms of the portfolio
risk and return.

\begin{figure}[tb] \label{fig:F1}

\includegraphics[width=1.0\textwidth]{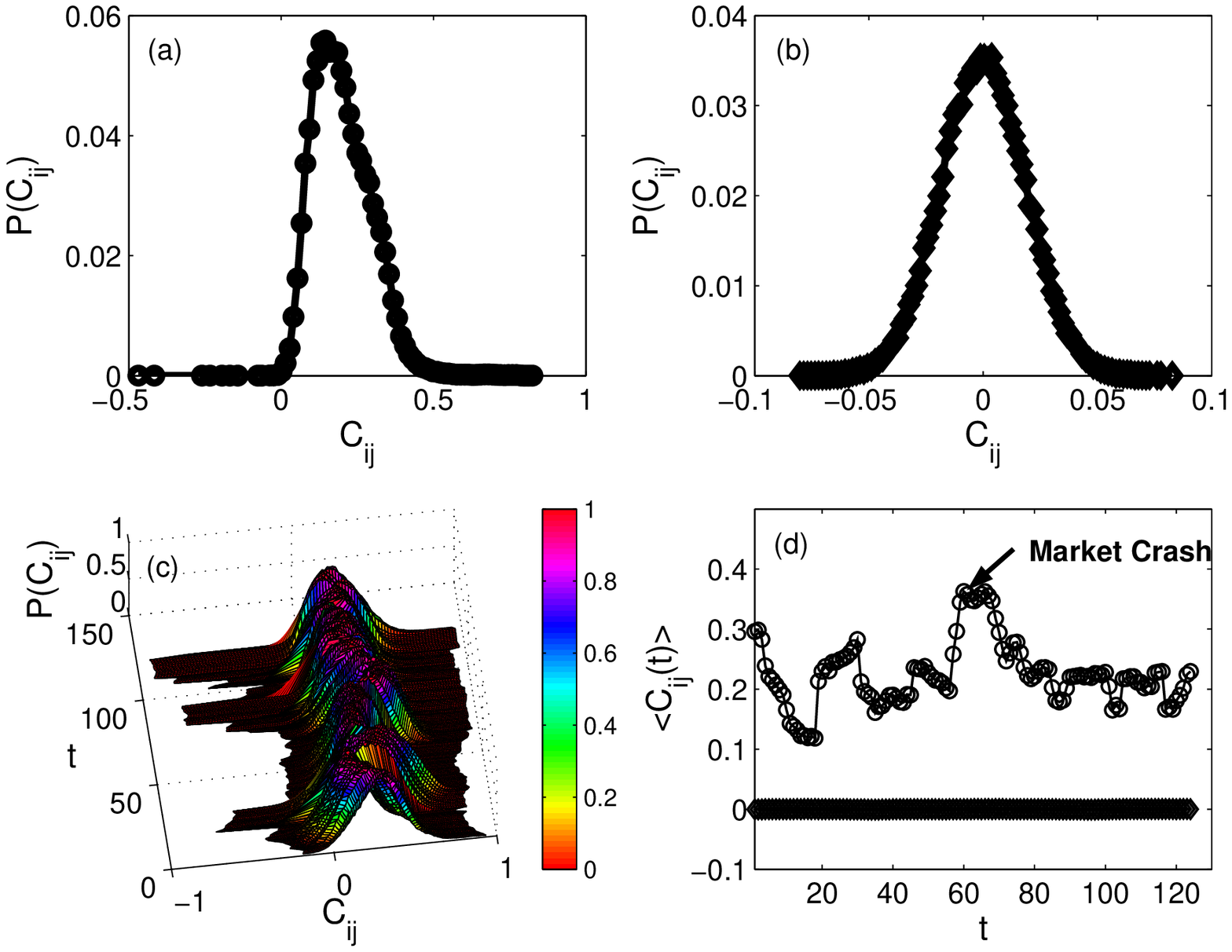}

\caption[0]{(a) and (b) show the distribution of the correlation
coefficients between stocks of 473 companies of taken from the
Korean stock market and random data, respectively. (c) displays
the distribution of the correlation matrices of the sub-periods by
shifting 21 days with 251 data points and (d) shows the average
values of each correlation matrix in (c)}
\end{figure}

\begin{figure}[tb] \label{fig.F1}

\includegraphics[width=1.0\textwidth]{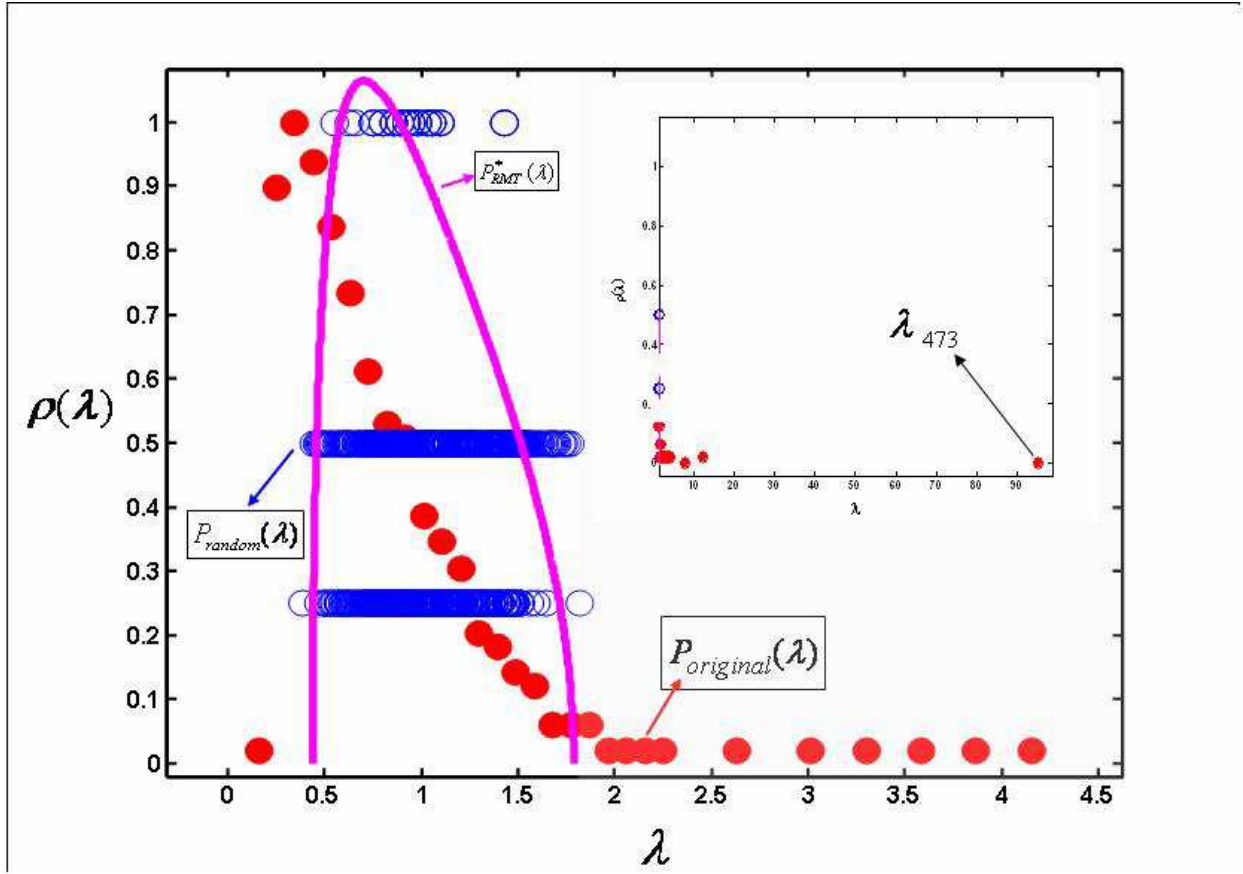}

\caption[0]{The distribution of the eigenvalues for correlation
matrix estimated using the 473 companies listed on the Korean
stock market, random data following the iid(0,1) process, and that
predicted by the RMT method. The red circles, blue circles, and
pink solid-line indicate the original time series, random data,
and theoretical lines, respectively.}
\end{figure}

\begin{figure}[tb] \label{F3}

\includegraphics[width=1.0\textwidth]{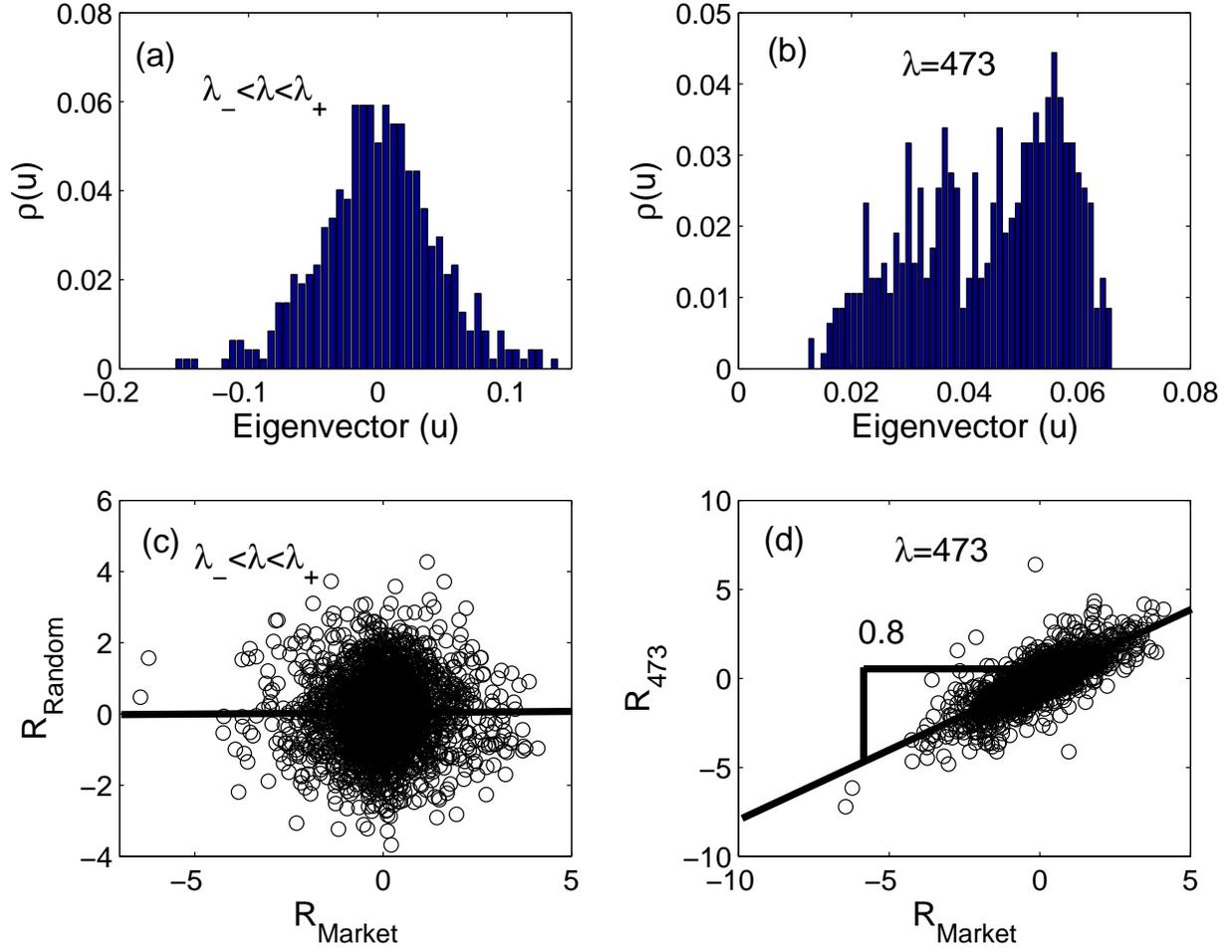}

\caption[0]{(a) and (b) show the distribution of both eigenvectors
corresponding to $\lambda_{100}$ and $\lambda_{473}$,
respectively. (c) and (d) display the $\beta$ coefficients between
the normalized market index and the time series created by
equation (9) for the eigenvalues $\lambda_{100}$ and
$\lambda_{473}$. The value of both $\beta_{100}$ and $\beta_{473}$
are zero and $0.8$, respectively.}
\end{figure}

\begin{figure}[tb] \label{F4}

\includegraphics[width=1.0\textwidth]{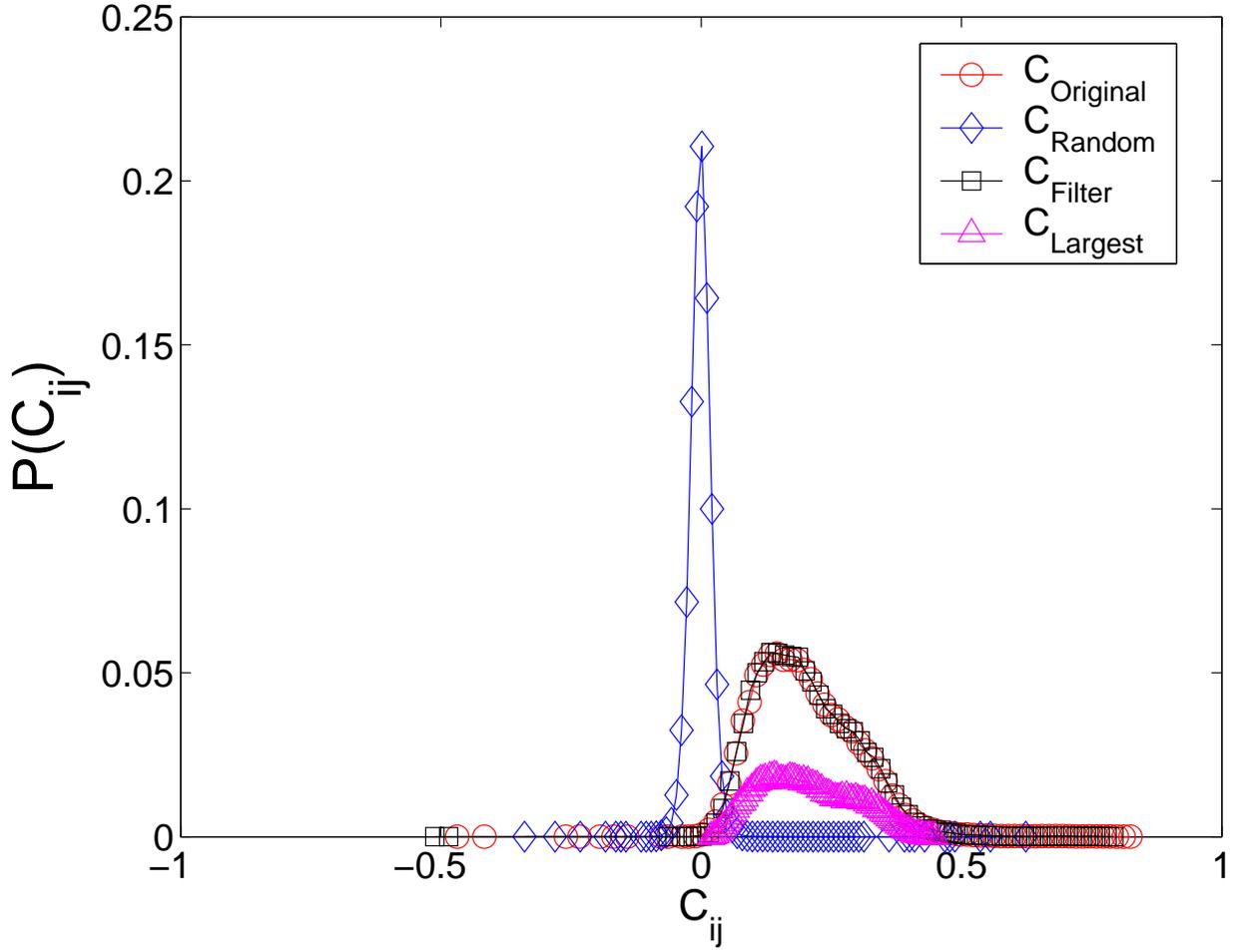}

\caption[0]{The distribution of the original correlation matrix,
$C_{original}$, and those created by the random matrix theory,
$C_{random}$, $C_{filter}$, and $C_{largest}$, respectively. The
red circles, blue diamonds, black squares and pink triangles
indicate the correlation matrices corresponding to the original,
random, filter and largest eigenvalue, respectively.}
\end{figure}

\begin{figure}[tb] \label{F5}

\includegraphics[width=1.0\textwidth]{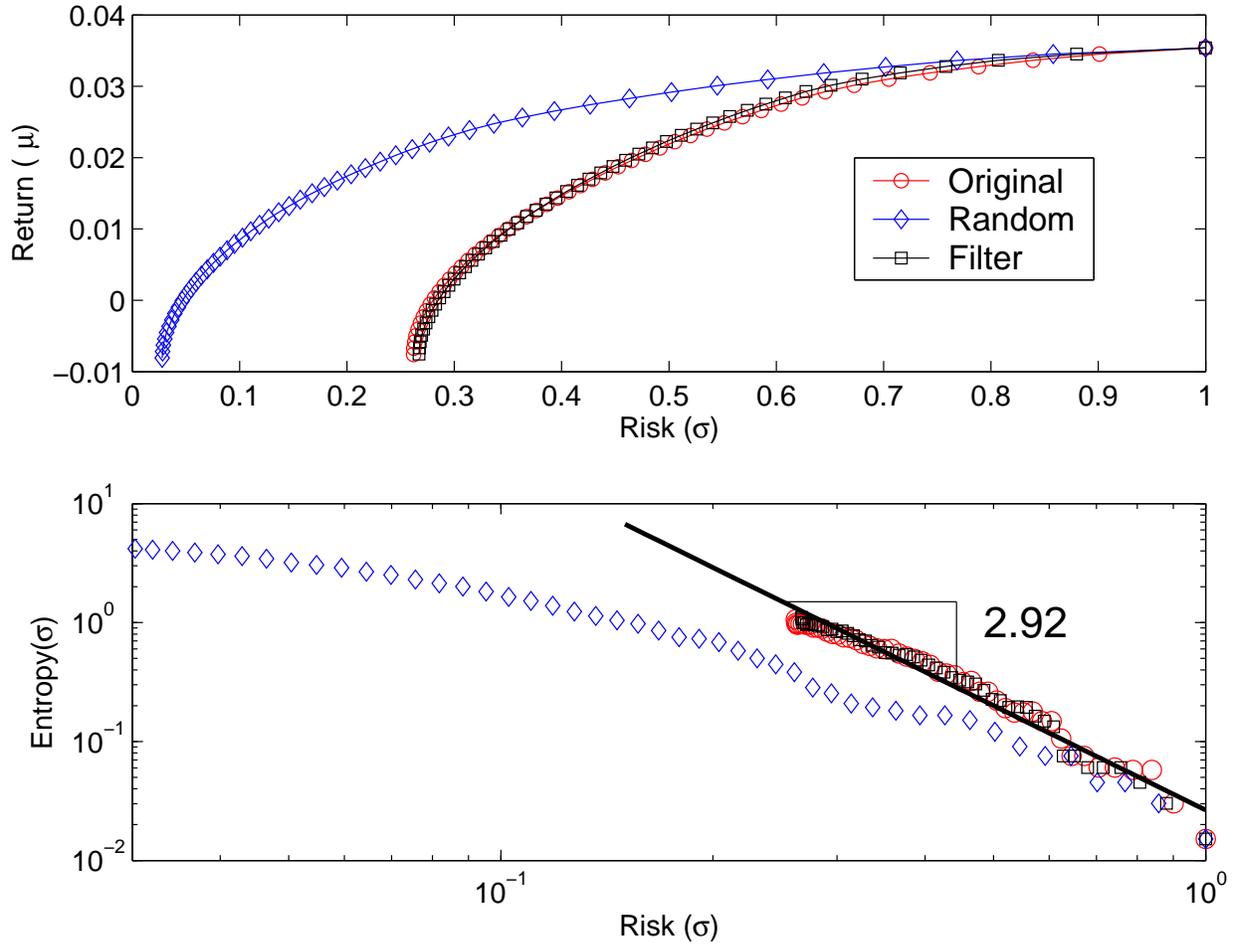}

\caption[0]{(a) shows the efficient portfolio frontier lines for
the original, $C_{original}$, random, $C_{random}$, and filter
correlation matrix,$C_{filter}$, respectively. (b) displays the
relationship between the entropy of the portfolio weights and the
portfolio risk. (c) displays the relationship between the entropy and the portfolio return.}
\end{figure}

\begin{figure}[tb] \label{F6}

\includegraphics[width=1.0\textwidth]{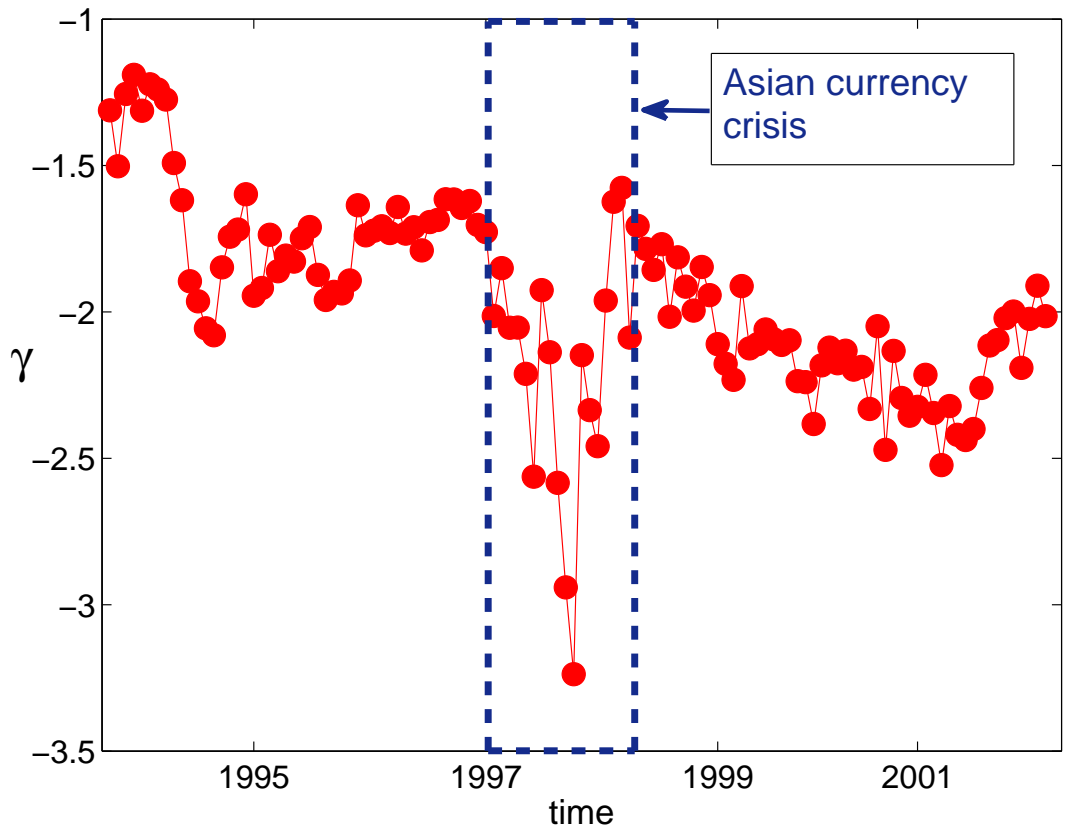}

\caption[0]{The exponent of power-law function, $E(\sigma) \sim \sigma^{\gamma}$ estimated by the relationship between the portfolio risks, $\sigma$, and the entropy of portfolio weights. }
\end{figure}

\end{document}